\documentclass[sigconf]{acmart}

\usepackage{pdfpages}

\AtBeginDocument{%
  }

\setcopyright{none}
\copyrightyear{}
\acmYear{}
\acmDOI{}

\acmConference[EduHPC-23 @ SC23]{EduHPC-23: Workshop on Education for High-Performance Computing -- Parallel Peachy Assigments}{Nov. 13, 2023}{Denver, CO}

\acmPrice{}
\acmISBN{}

\begin{document}

\title{Parallelizing a 1-Dim Nagel-Schreckenberg Traffic Model}


\author{Ramses van Zon}
\authornote{Both authors contributed equally to this research.}
\email{rzon@scinet.utoronto.ca}
\affiliation{%
  \institution{SciNet HPC Consortium, University of Toronto}
  \streetaddress{661 University Ave, suite 1140}
  \city{Toronto}
  \state{Ontario}
  \postcode{M5G 1M1}
  \country{Canada}
}

\author{Marcelo Ponce}
\authornotemark[1]
\email{m.ponce@utoronto.ca}
\orcid{1234-5678-9012}
\affiliation{%
  \institution{Department of Computer and Mathematical Sciences, University of Toronto Scarborough}
  \streetaddress{1265 Military Trail}
  \city{Toronto}
  \state{Ontario}
  \postcode{M1C 1A4}
  \country{Canada}
}



\begin{abstract}
The Nagel-Schreckenberg model is a stochastic one-dimensional traffic model
\cite{nagel1992cellular}.  In
this assignment, we guide students through the process of implementing
a shared-memory parallel and reproducible version of an existing
serial code that implements this model, and to analyze its scaling
behavior.

One of the key elements in this traffic model is the presence of
randomness, without which it would lack realistic phenomena such as
traffic jams.  Its implementation thus requires
techniques associated with \textit{Monte Carlo} simulations and
\textit{pseudo-random number generation} (PRNG).  PRNGs are
notoriously tricky to deal with in parallel when combined with the
requirement of reproducibility.

This assignment was created for the graduate course \textit{PHY1610
Scientific Computing for Physicists} at the University of
Toronto, which had its origin in the training program of the
SciNet HPC Consortium, and is also very suitable for other
scientific disciplines. Several variations of the assignment have been used over the years.

\end{abstract}





\maketitle





\section{Rationale}
The main rationale of this assignment is to present students with a
time-stepping, stochastic simulation and guide them through the
process of creating
a parallel implementation.
In this case, the system to simulate is the one-dimensional
Nagel-Schreckenberg traffic model\cite{nagel1992cellular}.  Simulating
this model requires using pseudo-random number generators\cite{Pressetal2007numrecipes} in parallel,
a tricky and often overlooked topic in scientific computing courses. 

We created this assignment for the ``Scientific Computing'' Physics
course taught to graduate students at the University of Toronto,
Canada.  This course aims to teach students programming skills to
develop scientific applications, using C/C++, best practices in
software engineering, use of well established libraries, and train
them in parallel computing techniques such as shared-memory
programming (i.e. OpenMP) and distributed-memory programming
(i.e. MPI).  The course, which consistently
gets positive course evaluations, is highly practical and applied, requiring
students to develop code on our teaching
cluster\footnote{\url{https://docs.scinet.utoronto.ca/index.php/Teach}}.
This course originated in the training program of the SciNet HPC
Consortium\footnote{\url{https://scinet.courses}}.  Because of this, the course is also
suitable for other scientific disciplines and many of its topics 
also fit in an undergraduate curriculum.


\begin{figure*}
  \includegraphics[width=\textwidth]{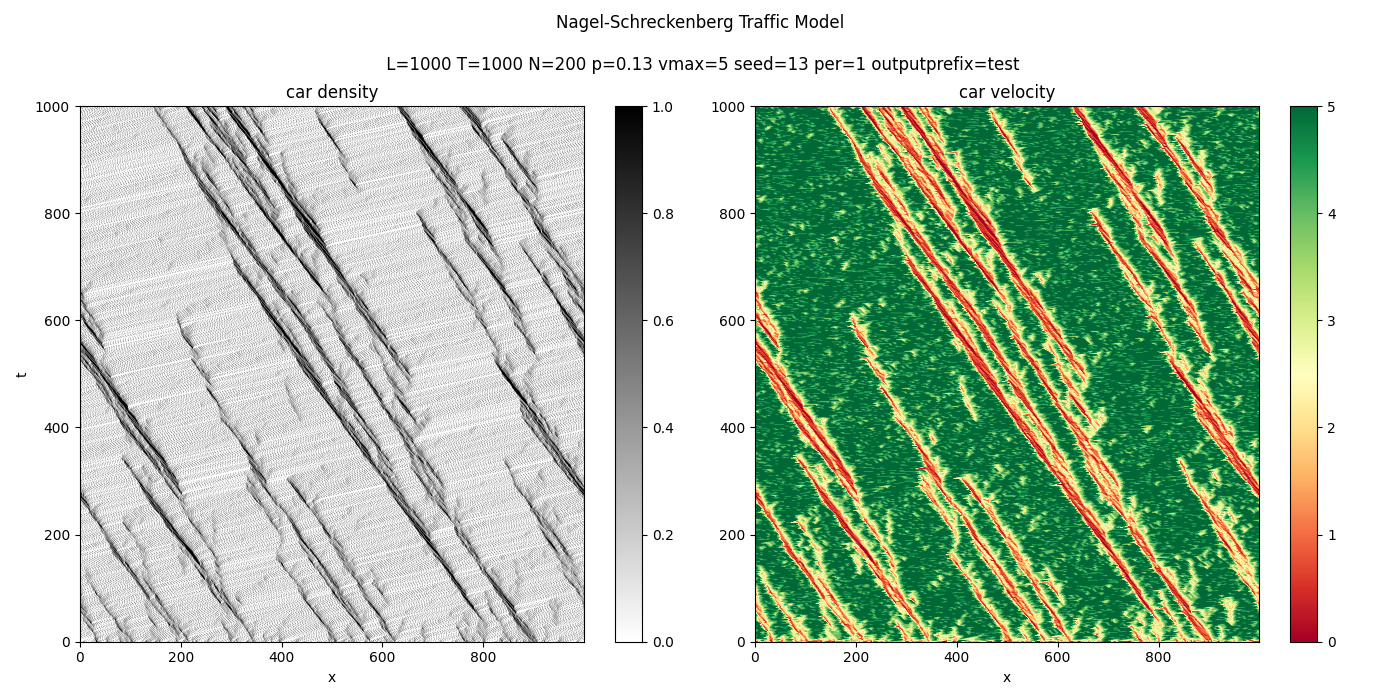}
  \caption{Visualization of the 1-dim simulation of the Nagel-Schreckenberg traffic model -- 
                with 200 cars, length of 1000, probability $p=0.13$ and maximum velocity $v_{max}=5$.
		The figure shows the emergence of irregularities ("traffic jams") in the flow of the vehicles,
		and how it propagates backwards in position and forward in time.
		Without the random contributions to the model, these irregularities would not occur.
	}
  \label{fig:NS_1dim_traffic_model}
\end{figure*}

\section{Concepts Covered}

The implementation of the Nagel-Schreckenberg traffic model
requires \textit{Monte Carlo} techniques, which in turn require a
\textit{pseudo-random number generator} (PRNG). Both of these are topics
covered in our course, and we use this model because it is an excellent and easily relatable example
of a stochastic simulation.  For the assignment, a starter code in \textit{C++}
is given, and \textit{OpenMP} should be used to parallelize the code on
\textit{shared-memory multi-core computers.}

One of the nice features of this problem is that it can be solved
using either a \textit{grid} representation or an \textit{agent-based} one.
The grid representation assigns a value to every point on the
circular road of
length $L$, while the agent-based implementation keeps track of the positions
and velocities of the $N$ cars as (two) vectors of length $N$ on that
circular road.
Each implementation has its advantages and disadvantages, but in particular
the agent-based approach significantly simplifies the parallelization of PRNG.

Pseudo-random numbers are generated by sequentially deriving a number from an
internal state that gets updated with every next number.
Before drawing the first number, the state is initialized from a `seed' value,
which is often a single integer.
The state update algorithm is deterministic, and therefore the sequence is
reproducible if the same seed is used.
The resulting sequence of numbers should nonetheless be nearly indistinguishable
from being statistically independent and evenly distributed. 

The circular road is also an example of \textit{periodic boundary conditions},
which have a wider applicability in computer simulations.

In the course, students are made familiar with programming in C++, best
practices in software development such as modularity, version control,
unit testing, documentation, use of external libraries, make, file
formats such as ASCII, binary, self-describing formats, etc.  To be
able to do this assignment, students should already have good working knowledge of
C++ and how to use the C++11 standard random library.  The starting
code is moderately modular, so familiarity with the concepts of C++
headers and implementation files is helpful.  Knowledge of OpenMP is required to do the assignment, including the \texttt{parallel}, \texttt{for}, and \texttt{threadprivate} compiler directives. The assignment was
designed and tested using the \emph{make} utility and a simple makefile for building and running the
software.  While we strongly recommend using this approach, it is
possible to use e.g. cmake as well, or to compile and run the code
manually. The code does not require external libraries.

One of the trickiest parts in the parallel implementation of this
model, and the one highlighted in this assignment, is dealing
with the PRNG in parallel in such a way that the output of the
parallel code \textit{exactly reproduces} that of the serial code.
Scientific \textbf{reproducibility} is a very urgent and critical
topic nowadays in many scientific disciplines that heavily rely on
computational technologies.  Without this requirement, one possible
solution to parallelize the code and its PRNG function, would be to have
each of the threads sampling from its own random number generator,
starting from different seeds, thus having a different random number
sequence in each thread.  However, this leads to different
results when the number of threads used changes.  Although this may be
tolerable in some situations, reproducibility between using various number of threads is
a requirement of this assignment.

Reproducibility requires there to be only one sequence of random numbers from
which to sample and shared among the different threads, so that
one gets the same results on the same
hardware, independent of the number of threads.
While generating a random number sequence is generally a serial process and therefore not parallelizable, for
several random number generators, there are algorithms for quickly ``moving ahead''.
Because these are not yet implemented
in the C++ standard random library, the starting code of this
assignment provides an implementation of this fast-forward algorithm
for one of the C++ linearly congruent generators.

\section{Limitations}
Depending on the parameters,
software implementation and characteristics of the hardware, the amount of computation can result in similar
or even smaller than the cost of I/O operations -- e.g. if we decide to save
data at higher rates of iterations.
We provide examples of parameter files to help to emphasize
the relevance of the computational part of the simulation, and it is possible to switch off output completely as well.

The scaling behavior that students may observe depends highly on the level to which they managed to reduce the cost of fast-forwarding the random number generators and other serial parts of the code.  Scaling beyond a single socket can be less than ideal due to NUMA effects. Finally, using more virtual cores than physical ones on CPUs which support ``simultaneous multithreading'' (also known as ``hyperthreading'') should be avoided; even if there is a small benefit, the timing results would be hard to interpret.

\section{Variations}

In this assignment, we focus on the parallelization of the algorithm,
in particular the PRNG and parallelization aspects of its
implementation using a shared-memory approach such as OpenMP.  In other
variation that we have used in the past, we have asked students to
create their own serial implementation from scratch, or to adapt the
output to use the NetCDF library.  This problem offers many other
opportunities for variation that address other HPC aspects.  One could
ask students to link to another PRNG library, to implement a
distributed-memory parallel code using the Message Passing Interface,
to port the code to use Graphics Processing Units, to run a series of
parameter study cases and take advantage of embarrassingly parallel
jobs, to perform scaling analysis, to do a performance analysis by
profiling the code, to change boundary conditions, etc.



\bibliographystyle{ACM-Reference-Format}
\bibliography{refs}

\appendix


\section{Online Resources}
\label{sec:ap_OnlineResources}

\begin{itemize}
	\item Course Materials from our "Scientific Computing for Physicists (Winter 2021)" course:
		\begin{itemize}
			\item Course Website:
				\\
				\url{https://education.scinet.utoronto.ca/course/view.php?id=1155}

			\item Assignment Part I: Serial Model
				\\
				\url{https://education.scinet.utoronto.ca/mod/assign/view.php?id=2605}

			\item Assignment Part II: Parallel Version
				\\
				\url{https://education.scinet.utoronto.ca/mod/assign/view.php?id=2606}
		\end{itemize}

	\item Repository, containing
		assignment handout
		and
		starter code:
		\begin{itemize}

			\item[] \url{https://github.com/Practical-Scientific-and-HPC-Computing/Traffic_EduHPC-23}
		\end{itemize}
\end{itemize}




\newpage

\section{Handout}
The following pages contain the assignment's handout, which is also accessible
from the public GitHub repository,
	\url{https://github.com/Practical-Scientific-and-HPC-Computing/Traffic_EduHPC-23}




\section*{Supplementary material}

# Parallelizing the 1-Dim Nagel-Schreckenberg Traffic Model

## The Nagel-Schreckenberg traffic model

The Nagel-Schreckenberg traffic model (Nagel and Schreckenberg 1992) is a 1D toy model used to generate traffic-like behaviour. This model is simple enough to avoid getting distracted with the details of its abstraction but to nonetheless demonstrate interesting emerging properties, such as congestion or stagnation points.

To describe the details of this model we will use dimensionless quantities, meaning that position, velocity and time, all have the same "units".

The model in this case is 1-dimensional (1-D), which means that there is only one spatial coordinate, which denotes the position of the cars along it. In the model, each car has a position $x$ and a velocity $v$ which evolve in time as follows:

1. If the velocity $v$ is below a maximum velocity, $v_{max}$, then increase $v$ by 1 (try to speed up).
2. If the car in front of the given car is at distance $d$ away, and $v \geq d$, then reduce $v$ to $d-1$ – we don't want to hit the car.
3. Add **randomness**: if $v > 0$ then with probability $p$ the car reduces its speed by 1.
4. The car moves ahead by $v$ steps (on a circular track).

These four rules can be summarized as,

$$\begin{cases} v & \leftarrow & \min(v+1, v_{max}) \\ v & \leftarrow & \min(v, d-1) \\ v & \leftarrow & (0, v-1) \text{ with probability } p \\ x & \leftarrow & x+v \end{cases}$$

where $\leftarrow$ means that the quantity on the left side of the arrow is updated with the prescription on the right side of it.

Hence, this model can be fully characterized by the following parameters:

- $v_{max}$, maximum allowed velocity;
- $N$, number of cars on the model;
- $L$, length of the road/track;
- $p$, threshold probability to decrease current velocity.

In case you are not familiar with using randomness – the ideas, concepts and algorithms are quite complex and involved, and a topic of ongoing research efforts. But in this particular case, it can be done in a relatively straightforward way:

- Draw a random number $r$ using a PRNG with uniform distribution on $[0, 1)$.
- For any chosen value $p \in [0, 1)$, the chance that $r$ is less than that value, is $p$ itself.
- So if $r$ is less than $p$, we will accept the move and decrease $v$ if possible.
- If $r$ is greater than or equal to $p$, we leave $v$ as it is, i.e., we reject the move.

It should be also noted that without the inclusion of *randomness* in the model (i.e. rule #3), the model would be deterministic, i.e. the cars will all evolve preserving the pattern as is set in the initial configuration.

Although this is a simple model, it has very rich dynamics and emerging properties, and has been used as foundations for more comprehensive models.

## Starter Code

For this assignment we provide a starter serial code, which can be found in the following repository:

https://github.com/Practical-Scientific-and-HPC-Computing/Traffic_EduHPC-23

The code simulates the dynamics of the Nagel-Schreckenberg traffic model for a set of cars that start from random positions on a road with periodic boundary conditions. This periodic boundary condition can be thought of as an approximation of a very long road, or as a circular road (i.e., a round-about). It implies that there is no incoming and outgoing motion of vehicles.

The code generates three files in the numpy array format, with their names given by the `outputprefix` parameter from the parameter file. For instance, if `outputprefix` is set to 'traffic', the files 'traffic-dens.npy', 'traffic-time.npy' and 'traffic-velo.npy', will contain the resulting positions, time and velocity of the cars for the simulation.

The code takes its simulation parameters from a parameter file which is specified as a command line argument. The parameter is a plain-text file containing the following parameters:

```
L = number of positions in the road (e.g. 500)
T = total number of time steps (e.g. 500)
N = number of cars in the road (e.g. 300)
p = probability of slowing down (e.g. 0.2)
vmax = maximum velocity (e.g. 2)
seed = random number seed (e.g. 13)
outputprefix = prefix for the name of the files to save the data (e.g. "traffic")
per = number of iterations how often output is done (e.g. 10)
```

The code is written in C/C++ using the `C++17` standard.

For Pseudo-Random Number Generation (PRNG) it uses the 'random' library from the C++ standard library.

We also include a python script `npyplot.py` to visualize the results from the simulation, which will a generate an image similar to the one shown in Fig.1.

The starter code is located in the `code` directory of this repository. The code is written in a modular fashion, with well defined modules to do specific tasks.

This the full list of files from the `Code` directory:

| Filename | Description |
| --- | --- |
| `Makefile` | make file |
| `CMakeLists.txt` | cmake file |
| `README.md` | README file |
| `test.ini` | parameter file for testing and reference case |
| `largetest.ini` | parameter file for longer and more demanding runs |
| `reference-test.png` | image with results from the reference case |
| `npyplot.py` | python script for visualizing results from the simulation |
| `npywriter.cpp` | writing routine module |
| `npywriter.h` | corresponding header file |
| `parameters.cpp` | parsing and processing of parameters module |
| `parameters.h` | corresponding header file |
| `traffic.cpp` | main driver function |

Most of the relevant code and functions for the assignment (with the exception of the IO routines) are in the `traffic.cpp` file.

**How to compile and run the code**

You need a relatively recent `g++` compiler (version 9 or above), `make` or `cmake` for compilation, and a `python` 3 installation with `numpy`, `matplotlib`, and `plotly` version 5 or above, to use the `npyplot.py` script for generating PNG files and generating interactive and HTML visualizations.

To compile the code, type

```
make
```

The `Makefile` also contains commands to run and plot results.

To run the smaller test case and plot the results in the form an png files, use

```
make test
```

To run the larger case and plot the results in the form an png files, use

```
make largetest
```

The parameters are set in `.ini` files that can be given as command line arguments, e.g.:

```
./traffic test.ini
```

Output is in 3 files, two with postfixes `-dens.npy`, `-velo.npy` that contain the density and velocity of the traffic on a grid at the times stored in the file with postfix `-time.npy`. These are in the binary numpy array format, which can be read e.g., by the `npyplot.py` script. This scripts needs the same parameter file as the simulation executable, so it can identify the resulting files and parameters used in the run, e.g.

```
python3 npyplot.py test.ini
```

The `npyplot.py` script can generate different formats for the plots. Using a second optional argument to the script, one could generate PNG (default or `png`), interactive in browser (`interactive`) or HTML (`html`) formats. E.g.

```
python3 npyplot.py test.ini interactive
```

**WARNING** the `interactive` visualization is a *beta* feature and may not work as expected with large datasets.

---

**IMPORTANT**

You should carefully look at the code, it is properly commented and documented. You must understand the code as many details of the implementation are critical for solving the assignment.

You should note that this is an *agent-based* implementation, i.e., an implementation that keeps track of the positions and velocities of the $N$ cars as (two) vectors of length $N$. An alternative representation, the *grid* representation, which assigns a value to every point on the road of length $L$ would be another possible way to tackle this problem, here is only used in the output routines.

Pay utmost attention to the random number generation. In our implementation, in spite of some of the short coming and disadvantages it may have compared to other PRNG methods, we use a *multiplicative linear congruent* generator, due to its ease of implementing a better discard method.

---

# THE ASSIGNMENT: Parallelize the 1-D Nagel-Schreckenberg Traffic Model

The aim of this assignment is to parallelize the Nagel-Schreckenberg model, for which we will use the starter code described before.

Your task is as follows:

1. Speed up this code using **shared-memory parallelism**, by adding OpenMP directives and thread-safe mechanisms for the pseudo-random number generator.

   Remember to always use `default(none)` in `pragma omp parallel` directives, as well as include the appropriate compilation flags.

   You can start with the often-used approach of having a different seed for the random number generator of each thread, but this will break reproducibility across different number of threads.

   Then, you must pay particular attention to how to treat and deal with the PRNG, specifically to guarantee the **reproducibility** of the results for the different implementations. I.e. as the *seed* of the PRNG engine is set in the parameter file, it is expected that for a given seed the results of serial and parallel runs should match when running on the same hardware independent of the number of threads used.

   To assess reproducibility, we suggest to use the `test.ini` parameters and visualize the results with `python npyplot.py test.ini`.

2. Run a strong scaling analysis for your implementation for the parameters in the `largetest.ini` file for increasing numbers of threads. Set `per=0` in the ini file to focus on how the computation scales. If you are running on a HPC system (e.g. supercomputer or cluster), create a submission script to submit your runs as jobs and automate the timing measurements.

3. Make a plot of the speedups as a function of the number of threads. Check if it is possible to estimate the serial fraction $f$ using Amdahl's law.

4. Create a weak scaling analysis by creating several cases of increasing values of `L` and `N` that are run with proportionally increasing number of threads. Create a plot of their runtime vs. number of threads.

5. Write a report based on the outcomes of the scaling analysis and the results from your previous estimate.

Optional:

6. Do a performance analysis to determine which parts of the program are the most expensive ones.

7. Focus on the I/O functions and look for opportunities for improvement by parallelizing them. Repeat the scaling analysis now with output enabled.

---

Nagel, Kai, and Michael Schreckenberg. 1992. "A Cellular Automaton Model for Freeway Traffic." *Journal de Physique I* 2 (12): 2221–9.

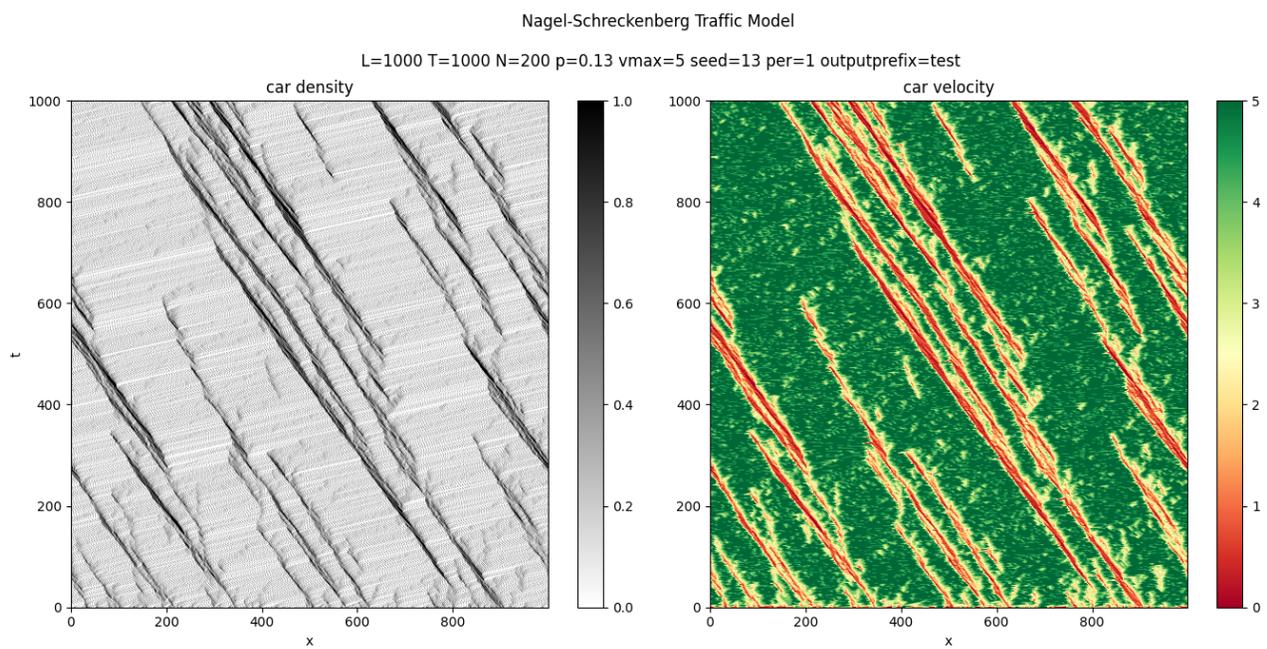

Figure 1: Visualization of the reference **test case** included with the starter code. Plots generated using the `npyplot.py` script.





\end{document}